          [2001/04/25 1.1 (PWD)]
\documentclass[a4paper,twocolumn]{esapub} 
\usepackage{natbib}
\usepackage{graphicx}
\def\simge{\mathrel{%
   \rlap{\raise 0.511ex \hbox{$>$}}{\lower 0.511ex \hbox{$\sim$}}}}
\def\simle{\mathrel{
   \rlap{\raise 0.511ex \hbox{$<$}}{\lower 0.511ex \hbox{$\sim$}}}}
\long\def\unmarkedfootnote#1{{\long\def\@makefntext##1{##1}\footnotetext{#1}}}

\title{INTEGRAL AND LIGHT DARK MATTER}
\author{M. Cass\'e$^{0,1,3}$}
\author{P. Fayet$^{2}$}
\author{S. Schanne$^{1}$}
\author{B. Cordier$^{1}$}
\author{J. Paul$^{1,4}$}
\affil{$^{1}$ CEA-Saclay, DSM/DAPNIA/Service d'Astrophysique, 91191 Gif-sur-Yvette, France}
\affil{$^{2}$ Laboratoire de Physique Th\'eorique de l'ENS
(CNRS),  24 rue Lhomond, 75231 Paris Cedex 05, France}

\begin{document}


\maketitle

\begin{abstract}
The\unmarkedfootnote{Corresponding author: m.casse@cea.fr \\
$^{3}$ Institut d'Astrophysique de Paris, 98 bis Boulevard Arago, 75014 Paris, France \\
$^{4}$ F\'ed\'eration de Recherche Astroparticule et Cosmologie, Coll\`ege de France, 11 Place Marcellin Berthelot, 75231 Paris, France
} nature of Dark Matter remains one of the outstanding questions of modern astrophysics. The success of the Cold Dark Matter
cosmological model argues strongly in favor of a major component of the dark matter being in the form of elementary particles, not yet discovered. Based on earlier theoretical
considerations, a possible link between the recent SPI/INTEGRAL measurement of an intense and extended emission of 511 keV photons (the hallmark of positron annihilation)
from the central Galaxy, and this mysterious component of the Universe, has been established advocating the existence of a {\sl light} dark matter (LDM) particle (at variance with the neutralino, in general considered as very {\sl heavy}).

\vspace{-2mm}

We show that it can explain the 511 keV emission mapped with SPI/INTEGRAL without overproducing undesirable signals like
high energy gamma-rays arising from $\pi^\circ$ decays, and radio synchrotron photons emitted by high energy positrons circulating in magnetic fields.
Combining the annihilation line constraint with the cosmological one (i.e. that the relic LDM energy density reaches about 23\% of the density of the Universe), one can restrict the main properties of the light dark matter
particle.
Its mass should lie between $\,\approx$ 1 and 100 MeV, and the required annihilation cross section, velocity dependent, should be significantly larger
than for weak interactions, and may be induced by the virtual production of
a new light neutral spin 1 boson $U$.

\vspace{-2mm}

On astrophysical grounds, the best target to validate the LDM proposal seems to be the observation by SPI/INTEGRAL and future gamma ray telescopes of the annihilation line from the Sagittarius dwarf galaxy and the Palomar-13 globular cluster, thought to be dominated by dark matter.

\vspace{-1mm}
\end{abstract}

\section{Introduction}

In the currently popular Cold Dark Matter (CDM) cosmological model with a cosmological constant $\Lambda$, the Universe is composed of
$\,\approx 4\,\%$ baryonic matter and $\approx 23\,\%$ cold, collision-less dark matter, and is made flat by a cosmological constant \citep{spergel2003}. The very nature of dark matter is one of the tantalizing and yet unanswered questions of contemporary astrophysics. While its existence has been inferred for several decades, its amount has been clearly assessed on various astrophysical scales only in the recent past. The timed questions are its distribution and above all its true nature. The precise knowledge of its density profile is likely to be decisive in answering the ultimate question of the identification of the kind of particle(s) composing dark matter. The success of the CDM cosmology, combined with dark energy ($\Lambda$), argues strongly for a major component of the dark matter being in the form of fundamental (supersymmetric?) particles.

One possible way of probing the nature of dark matter particles is to look for their annihilation signal (annihilation of dark matter particles on their anti-particles, knowing that both could be identical). The best regions to search for this annihilation radiation is where dark matter accumulates, i.e. in deep gravitational potential wells. The annihilation rate is proportional to the square of the number density of the dark matter particles (in the case where $dm=\overline{dm}$). The Galactic Center and its surroundings are therefore obvious targets because of their density and proximity.

A wide literature exists on the prospects of observing dark-matter-induced gamma rays from the Galactic Center and clumps of dark matter in the Galactic halo (\cite{bertone2002} and references therein).
But these gamma rays are at high energy, outside the INTEGRAL grasp, and are prime targets for high energy gamma ray satellite experiments (like GLAST), and atmospheric Cherenkov telescopes (such as HESS, VERITAS, MAGIC and CANGAROO). In the following, however, we advocate a link between dark matter and
{\it \,low energy\,} gamma rays, within the reach of INTEGRAL.

511 keV photons are the signature of electron-positron annihilations. Recent observations of 511 keV photons from the central region of the Galaxy, performed by the SPI spectrometer on board of the INTEGRAL satellite \citep{ jean2003,knodlseder2003,jean2004,weidenspointner2004} confirm previously reported observations \citep{leventhal1993, milne2002} and improve the spatial resolution. The SPI/INTEGRAL data indicate that the 511 keV emission zone is spherically symmetric, best fitted by a gaussian distribution with 9$^\circ$ (FWHM), and inconsistent with a single point source
(although it is premature to exclude multiple point sources).
The large extension of the emission region around the Galactic Center combined with the surprising failure of the leading candidates (SN~Ia), as shown in \cite{schanne2004}, opens the way to new speculations. A possible link between the 511 keV emission line mapped by SPI/INTEGRAL and a new kind of dark matter, bosonic and light, that had been discussed previously \citep{boehm2003a,boehm2003b} was proposed by \cite{boehm2004a}, since other, more conventional, explanations are questionable.

\section{The failure of SN~Ia and difficulties of hypernovae}

The story begins with the fact that the most generous positron  sources proposed up to now,
supernovae SN~Ia, fall short of explaining the intensity of the measured 511 keV emission line
from the central part of the Galaxy, reminiscent from the Galactic bulge. Strangely enough, this point has remained unnoticed until recently \citep{casse2004}, whereas it is already apparent in the OSSE/GRO data. Indeed the Galactic bulge is like a tiny elliptical galaxy embedded in a spiral. Thus scaling the supernova rate to that of elliptical galaxies, as a function of the mass  \citep{cappellaro2003}, we get a small explosion rate, $\,\approx 0.03$ SN~Ia per century \citep{schanne2004}, insufficient to explain the required positron injection rate of about 10$^{43}$ s$^{-1}$, given the low escape fraction of positrons from SN~Ia \citep{milne2002}.

The striking roundness of the emission, reminiscent of the Galactic bulge, leads naturally to propose a whole population of unresolved sources lying in the Galactic bulge, i.e. old and small mass stars (single or in binary systems), or stellar remnants. But one cannot exclude that a recent event like SN2003dh linked to the young population sitting in the Central Molecular Bulge, a region of active star formation \citep{figer2004}, has injected a large amount of positrons that have invaded the bulge \citep{parizot2004} with the help of oriented and ordered magnetic fields, non thermal filaments \citep{yusef-zadeh2004} and a bipolar galactic wind \citep{bland-hawthorn2003}. The rate of hypernovae \citep{podsiadlowski2004} seems insufficient to fulfil the positron constraint, however advocating the existence of a stellar burst in the last million years in the Galactic Center able to trigger the bipolar galactic wind makes the situation less desperate \citep{schanne2004}.
Thus $\beta^+$ radioactivity from freshly synthesized nuclei remains a possibility but still fragile.
Other potential sources of positrons abound, but they appear considerably weaker than SN~Ia and hypernovae of the SN2003dh type: flare stars, low mass X ray binaries (LMXRB) \citep{prantzos2004}, more particularly micro-quasars. In general all of these candidates remain problematic since the required injection rate of positrons in the Galactic bulge as deduced from the SPI/INTEGRAL measurement is impressively high compared to their possibilities. Moreover their distributions do not fit with the 511 keV photon one \citep{paul2004}.

\vspace{-.4mm}

In the absence of a solid astrophysical explanation it is worth considering that the galactic 511 keV line emission may be of a radically different origin (i.e. not of nuclear origin), opening the door to new physics, independently of supersymmetry (or possibly beyond it).

\vspace{-.3mm}

\section{Dismiss of heavy neutralinos}

\vspace{-1.5mm}

The foremost candidate for the cold dark matter composing galactic halos is the lightest
supersymmetric particle, in general a neutralino $\chi$ (combination of neutral gauginos and higgsinos),
protected from decay by $R$-parity conservation in supersymmetric extensions of the Standard Model \citep{fayet1977,farrar1978}.
As such, the neutralino is stable and can serve as a Weakly Interacting Massive Particle, or WIMP.
The relic density of neutralinos can be identified with the one deduced from observations (WMAP),
given that they are weakly coupled to matter,
and that their mass is expected to be very large.
In fact, collider experiments now constrain $m_\chi \simge $ 50 GeV, although this lower bound depends
on some specific hypothesis.
Obtaining the appropriate relic density of dark matter particles also requires, in general,
$m_\chi \simle$ 500 GeV at most (excepted for narrow specific regions in the parameter space).

\vspace{-.5mm}

These particles, being of the Majorana type, are their own antiparticles, and they can annihilate in the central region of the Galaxy, where they accumulate. Their annihilations proceed in various ways, depending on their mass (and on the parameters of the specific models considered).
In addition to the direct $\,e^+e^-$ annihilation channel, there are many other possible channels that
ultimately lead also to positrons through cascade decays \citep{baltz2001,peirani2004,prada2004}.
These annihilation reactions depend on a number of free parameters of the theory, at least five in the simplest case (and in general significantly more).

\vspace{-.5mm}

Positrons could be produced by neutralino annihilations in the central region of the Galaxy. In this case the observed 511 keV emission could be the result of a double annihilation
process $\,\chi\, \chi \to e^+\,+\,...\ , $ \,followed by $e^+ e^- \to \gamma \,\gamma\,$.
Neutralino pair annihilations may lead to the emission of gamma rays and high energy positrons through various decay channels which may imply the production of $W^\pm$ and $Z$ bosons, as well as
quark anti-quark pairs. This leads in particular to the production of neutral and charged pions,
decaying eventually into photons, electrons and positrons, as follows:

$\ \ \ \ \ \ \ \ \chi\, \chi\ \ \to \ \ W^+ W^-\ $ (\,or $\,Z \,Z \,$)\ \ ,\ \ ...\ \ ,

$\ \ \ \ \ \ \ \ \,q \ \bar{q} \ \ \to \ \ \pi\ +\ ... \ \,,$

\ \ \ \ $\left\{ \begin{array}{lcl}
\pi^0 &\to& \ \gamma\, \gamma \ \ \ , \vspace{2mm} \\
\pi^+ &\to& \ \mu^+\ \ \to \ \ e^+ \ \,,\vspace{2mm}\\
\pi^- &\to& \ \mu^- \ \ \to \ \ e^- \ \,.
\end{array} \right.$

Neutralino annihilations not only generate neutral pions which decay into $\gamma\gamma$, but also a
similar number of charged pions that decay ultimately into positrons and electrons,
the number of high energy photons per positron (or electron) produced being roughly
of the order of unity.
If one assumes that positrons in the central region of the Galaxy originate from neutralino annihilations, a large flux of high energy gamma rays would then ensue,
in strong contradiction with the EGRET upper limit.  Moreover, explaining the positron rate responsible for the 511 keV emission from the central region of the Galaxy in terms of neutralino annihilations
(at a rate of about 10$^{43}$ s$^{-1}$) would lead to a strong overestimate of the radio emission of this region in the presence of magnetic fields, the electrons and positrons produced generating excessive synchrotron radiation.

In addition, the positrons are emitted at high energy due to the large mass of the neutralino,
generally considered to be heavier than about 50 GeV.  Since positron annihilations should take place almost at rest, the positrons should lose almost all their kinetic energy before annihilating.  The
very large kinetic energy of such positrons (and similarly electrons, ...) should then be radiated (in various ways, synchrotron emission, bremsstrahlung, inverse Compton) and the whole region should shine brightly at different wavelengths, from radio to gamma rays, which is not observed \citep{paul2004}.
It seems thus that heavy neutralinos must be discarded as the positron source required by the INTEGRAL \,511 keV measurement.

However, the possibility that high energy gamma rays may be identified by forthcoming atmospheric Cherenkov telescopes (CANGOROO, VERITAS, HESS) or satellite borne detectors (like GLAST) excites considerable interest  \citep{taylor2003}. In view of this, it is interesting to note that theories inspired
from supersymmetry or extended supersymmetry may lead to {\it two} stable dark matter particles \citep{boehm2003b}, a light one that may be responsible
for the 511 keV line, and a heavy one for high energy gamma rays \citep{boehm2004b}.

\vspace{1cm}

\section{Light dark matter}

Weakly-interacting stable massive neutral particles, taken as possible dark matter candidates,
should not be too light, otherwise they would not have been able to annihilate sufficiently in the early Universe, and would therefore overclose it.
The best motivated candidate is the heavy neutralino,
considered in Section 3.
On the other hand, before the recent observations of the 511 keV gamma ray line from the Galactic Center, it was shown that light spin 0 dark matter particles could indeed exist \citep{boehm2003b}, as an alternative to the standard scenario of heavy WIMPs, and indicated that ``a gamma ray signature from the
galactic center at low energy could be due to the existence of a light new gauge boson'',
responsible for the annihilations of such light dark matter particles.

For light dark matter (LDM) particles to exist, we need new efficient mechanisms responsible for their pair annihilations -- otherwise they would normally, by far, overclose the Universe.
But on the other hand they should have no significant direct coupling to the $Z$ boson, otherwise they would have been produced copiously in $Z$ decays at LEP.
Still they would have to annihilate sufficiently, and in fact {\it much more strongly than through ordinary weak interactions}, otherwise their relic density would be too large!
How can this be possible\,?

These annihilations (at an appropriate rate) may indeed result from the virtual production of a new neutral spin 1 gauge boson $U,$ {\it \,light} but {\it very weakly coupled}, such as one introduced long ago, with the standard $\,SU(3)\times SU(2)\times U(1)\,$ gauge group of strong and electroweak interactions extended to include a new, spontaneously broken, extra $\,U(1)$ symmetry \citep{fayet1980,fayet1981,fayet1990}.
But on the other hand, a large annihilation cross section of LDM particles could lead to an excessive production of gamma rays from the Galactic Center, if $\,m_{dm}$ is less than about 100 MeV \citep{boehm2003a}, above what is really observed by the CGRO experiments OSSE, COMPTEL and EGRET \citep{hasinger1996,sreekumar1998}. This indicates that LDM particles should have a {\sl velocity dependent annihilation cross section}: the cross section for the
late annihilations of residual dark matter particles should be at least $\,\approx 10^5$, 10$^3$ or 10$^2$ times smaller than estimated at the freeze out epoch, for  $m_{dm}$ = 1, 10 or 100 MeV,
respectively.

It may indeed be possible to reconcile a sufficiently large annihilation cross section at freeze out (just after the big-bang), with an acceptable rate of gamma ray and positron production at present epoch, if the average of the product of the annihilation cross section with the velocity of LDM $\,<\!\sigma_{ann} \,v\!>\,$\, is proportional to $v^2$, automatically suppressing the late annihilation of relic dark matter particles \citep{boehm2003a}. This requirement is naturally satisfied for annihilations induced through the virtual production of a light $U$ boson, which shows
that the idea of light dark matter particles is indeed viable.
Other constraints from various particle physics experiments also appear compatible with this hypothesis, the smallness of the $U$ couplings to ordinary particles, as compared to
the electromagnetic coupling $e$, by several orders of magnitude, accounting for the fact that neither this $U$ boson, nor the light dark matter candidate \,-- which may have spin 0, or also spin
$\frac{1}{2}$ --\, has been observed yet \citep{boehm2003b,fayet2004}.
In addition, the $U$ boson should be {\it vectorially\,} coupled to ordinary matter particles,
so as to avoid a problematic axion-like behavior of its longitudinal polarization state.

Now, can the Galactic Center positron source be identified with the lower dark halo, occupied by LDM particles, which by annihilation would produce positrons?

\vspace{-1.5mm}
1. In order to have a 511 keV photon flux at the observed level, one needs a sufficient rate of positron annihilations in the central zone (while for a given dark matter density the number density of dark matter particles is inversely proportional to their mass).
This requires, in turn, significantly smaller masses than the $\,\sim$ GeV\,--\,TeV mass range preferred by most WIMP models.

\vspace{-2mm}

2. The produced $e^+$ must slow down to rest before annihilation.
In order to prevent leakage from the bulge region, the positrons produced should have rather low energy.
This again puts dark matter particles in the mass range of \,1\,--\,100 MeV \citep{boehm2004a}.

\vspace{-2mm}

3. There is, furthermore, a preference for the lower end of the above mass interval,
so as to avoid relatively energetic positrons and electrons from dark matter annihilations
to lead to too much energy dissipated in gamma rays, ...

Existing relic density constraints (e.g. $\!0.094 \,< \Omega_{dm} \,h^2 < $ 0.129) at 2$\sigma$ from WMAP \citep{spergel2003}, together with the required annihilation rate at the present time in the halo, can be recovered, provided the mean annihilation cross section times velocity ($<\!\sigma_{ann} v\!>$)
behaves as the square of the dark matter velocity.
Recent big-bang nucleosynthesis (BBN) calculations in presence of LDM particles have been performed, showing that these particles can lower the production of $^4$He and deuterium. For a mass between 4 and 10 MeV they even improve the agreement between the abundances predicted and the primordial ones derived from observations \citep{serpico2004}.

To summarize, a small mass together with a large annihilation cross section into $\,e^+ e^-$ pairs
\,-- as it can result from a new gauge interaction --\,
are required for such Light Dark Matter particles.
Other possible explanations in terms of decays of long-lived (quasi-stable) rather light relic particles (such as sterile neutrinos, or axinos with $R_p$-violating decays) into  $\,e^+ e^-$ pairs have also been proposed \citep{picciotto2004,hooper2004}.

\section{Dark matter density profile \hbox{and clumpiness}}

One of the keys of the problem is the distribution of the 511 keV emission projected on the sky. Indeed the emissivity depends both on particle physics parameters and on purely
astrophysical ones. Formally, the local emissivity depends on the square of the dark matter density modulated by the distribution of electrons on which positrons annihilate, integrated on the line of sight in various directions, and integrated again over the opening angle of the detector:

$\Phi_{511} ( r)\ \propto\ \rho_\chi ( r)^2 \  n_e (r )$\ .

It is clearly of importance to calculate the distribution of the annihilation signal and to compare it with the SPI/INTEGRAL observations. Unfortunately this is plagued with uncertainties. On one hand, the dark matter distribution is still highly debated and on the other hand, the gas content and distribution in the bulge are poorly known.
We then have to rely on purely theoretical arguments.

Within the framework of the $\Lambda$-CDM model, the growth of structures is seeded by a nearly scale-invariant spectrum of density fluctuations, supposedly generated during cosmological inflation. Structures form hierarchically, with small objects forming first and subsequently merging into larger structures over time. This widely accepted version of structure formation predicts the existence of a large number of self-bound substructures in the dark halo of our Galaxy \citep{moore1999,stoehr2003} that could give rise to an enhanced gamma ray signal due to annihilation of dark matter particles in the dense inner region (\cite{koushiappas2003} and references therein) and positron emission as well.

Different profiles for the dark matter energy density have been proposed, based on high resolution $N$-body simulations, differing vastly one from another
\citep{moore1999,persic1998,navarro1996,bahcall1984}.
If the dark matter density is cusped as $1/r$ at small radii, then the positron production,
and the gamma ray flux as well, would be much enhanced. The central massive black hole could also have an impact on the dark matter distribution concentrating it in the center of the Galaxy and producing a spike \citep{bertone2002}, but at the  modest spatial resolution of SPI, this extra concentration would have little effect.
Inspired by the persistence of substructures in numerical simulations, it has been argued that a substantial signal in the gamma ray observation (in our case the positron production) can be expected from such clumps (sub-halos) provided they are themselves cusped (\cite{evans2003} and references therein). These conclusions, however, have been contested as being unduly optimistic
\citep{stoehr2003}.
Indeed, according to the adopted profile, the numbers of detectable sub-halos differ dramatically \citep{koushiappas2003} since the luminosity, and therefore detectability of a sub-halo, is given by integration of the square of the dark matter density along the line of sight to the sub-halo.
More concretely, there is a substantial body of astrophysical evidence that the halo of our Galaxy is not cusped at all \citep{evans2003}. This is also in line with the interpretation of the 511 keV angular distribution in terms of dark matter annihilation, since using the simple power law parameterization of the dark matter energy density ($\rho_{dm}\sim r^{-\gamma}$ near the center), the range 0.4 $<\gamma<$ 0.8 is indicated \citep{boehm2004a}.
Indeed, one additional interest of the light dark matter scenario, if it turns out to be correct, is the possibility to probe the slope (steepness) of the dark halo profile in the central region of the Milky Way, and optimistically its clumpiness.

Summarizing, according to $N$-body simulations, dark substructures of masses higher than
$\,10^7\ M_\odot$ survive in significant numbers in the halo.
With more data coming in the future, with improved angular resolution, the 511 keV flux could provide more details on the distribution of dark matter in the inner Galaxy, provided the problem of the positron annihilation is clarified. The annihilation medium (except in the Central Molecular Zone) is very diluted. The positron annihilation rate is $\,R(r) = \,n_+ n_- <\sigma v>\, $,
\,where $n_+$ and $n_-$ stand as the positron and electron number densities, and $\sigma$ is the positron annihilation cross section, depending on the temperature and density of the ambient gas. Indeed many processes compete (\cite{ballmoos2003} and references therein) and the physics of annihilation is complex. Careful work is required to calculate the positron propagation and annihilation in the conditions of the Galactic bulge, and to deduce the dark matter distribution from that of the 511 keV annihilation line emission.

\section{Experimental tests}
\subsection{The Sagittarius dwarf galaxy}

Today, it seems difficult to explain the injection rate of positrons in the central Galaxy and the morphology of the 511 keV line emission observed by SPI/INTEGRAL on the basis of standard astrophysical scenarios. Thus it is worth pursuing the idea that the 511 keV emission is the signature of a rather exotic process, like for instance the annihilation of positrons induced by light dark matter particles
in the lower dark halo of our Galaxy. If it is the case, it should be an universal process, that should show up in favorable situations. In this line of thought, \cite{hooper2003} have proposed to test the light dark matter hypothesis on the Sagittarius Dwarf Galaxy (SDG), a close-by galaxy for which dark matter annihilations may be significant. A large dark matter annihilation rate is then predicted from this galaxy and as it contains comparatively few stars ($M/L$ = 21 in solar units), the detection of a substantial 511 keV emission line from SDG would provide a strong evidence for the light dark matter hypothesis.
The predicted flux at 511 keV from SDG, $(1 - 7)\times 10^{-4}$ ph cm$^{-2}$ s$^{-1}$, \,is a priori observable by SPI/INTEGRAL, but it is still undetected.

This estimate, however, relies on a number of unverified assumptions discussed in \cite{cordier2004},
the most important one being that annihilation of positrons effectively occurs in this environment
(i.e. that the gas density in all forms within SDG and/or its halo is sufficient to guarantee an annihilation time less than the age of this galaxy). This could be checked, for instance, by observations of the OVI absorption line in the UV by the FUSE satellite \citep{vidal2004}. Note also that the current mass estimate of SDG is in the range $(5-20) \times 10^8\ M_\odot$, moreover the tidal radius is ill defined in this system. Another drawback is that the SDG is in a state of tidal disruption, as evidenced by a trail of stars and gas escaping from its core \citep{cordier2004}, thus the emission should be more diffuse and therefore more difficult to detect than initially thought.

\subsection{Palomar-13}

Another interesting target for SPI/INTEGRAL is Palomar-13, a quite unusual globular cluster located at $\,(l,b)  =  \,(87.1^\circ,\ -\,42.7^\circ$) which, contrary to the general members of its class, could be exceptionally dark matter rich \citep{cote2002}. The luminosity of this remote (24.3 kpc) globular cluster, measured within its tidal radius (23 pc), is $\,2.8\,\times\,10^3 \,L_\odot$, corresponding to  $\,\approx 10^5 \,M_\odot$.
With a $M/L$ ratio of about 40, it could be a long sought dark matter dense clump, very favorable to dark matter research \citep{giraud2002}.
Since dark matter annihilation is a two body process, the luminosity of a given source scales as the integral along the line of sight of the dark matter density squared ($L\, \sim\rho^2R$), and the observed flux as $\Phi\sim M^2R^{-5} D^{-2}$. Applying this relationship to the Galactic Center region (located at $D=8$ kpc, the dark matter $M$ within a radius $R=1$ kpc depending on the radial distribution adopted) and to the globular cluster Palomar-13 (located at $D=24.3$ kpc, $M=1.2\times 10^5\, M_\odot$ within $R=23$ pc) we obtain the fluxes shown in Table \ref{tab:table1}.
Note that there is a possible drawback to the hypothesis of dark matter dominance. The high velocity dispersion of stars could be due to the dissolution of the cluster under the effect of tidal forces.

\begin{table}
  \begin{center}
    \caption{Prediction of the 511 keV brightness $\Phi$ of Palomar-13 for different dark matter distributions in our Galaxy
    \citep{moore1999,bahcall1984,navarro1996,persic1998}, where $M$ is the dark matter mass enclosed within the central $R=1$ kpc of our Galaxy.}
    \vspace{1em}
    \renewcommand{\arraystretch}{1.2}
    \begin{tabular}[h]{|l|r|l|}
      \hline && \\ [-3mm]
      &	\ $M/M_\odot$ \  &	$\ \Phi$ (cm$^{-2}$ s$^{-1}$)\ \\ && \\ [-3mm]
      \hline && \\ [-3mm]
\ Moore\  & \ 2.5 10$^9$ \ & \ \ \ \ 4\ \ 10$^{-5}$ \\
\ BS	& \ 8\ \ 10$^8$ \ & \ \ \ 3.9 10$^{-4}$ \\
\ NFW	& \ 3.8 10$^8$ \ & \ \ \ 1.7 10$^{-3}$ \\
\ PS	& \ 1.8 10$^8$ \ &\ \ \ 8.1 10$^{-2}$ \\ [-3mm] &&	\\ \hline
      \end{tabular}
    \label{tab:table1}
  \end{center}
\end{table}

Assuming total annihilation of the positrons released (i.e. no escape), we get different predictions according to the choice of the dark matter distribution \citep{moore1999,bahcall1984,navarro1996,persic1998}.
All but the \cite{moore1999} one give fluxes easily observable by SPI/INTEGRAL. However the validity
of the no escape condition of positrons is not guaranteed. A dedicated study of the matter content (supposedly rather high due to the fact that planetary ejecta do not escape from the deep  gravitational potential well) and magnetic field of Palomar-13 is mandatory, and above all a demonstration of the fact that it is not a dissolving cluster.


\subsection{Terrestrial searches}

An obvious constraint on LDM is that it has no significant direct coupling to the $Z$ boson,
otherwise it would have been discovered in $Z$ decays at LEP. This is not an unreasonable requirement, since other putative particles like right-handed neutrinos and binos, for instance, subscribe to it.
Again, the smallness of the $U$ coupling to ordinary matter as compared to the electromagnetic one
$e$  accounts for the fact that LDM particles have not been observed yet \citep{boehm2003b,fayet2004}.
Such particles would have escaped detection in $\,e^+e^-$ colliders as their cross section for
the production of anomalous single photons, in particular,
is below the sensitivity of past experiments. Note that while $U$ induced interactions
would generally be {\it stronger than weak interactions\,} at lower energies
(at which weak interactions are really weak!),
they would be much weaker at higher energies (for which weak interaction cross sections grow larger),
as the result of $\,U$ propagator effects.
$U$ exchanges could also lead to additional (positive) contributions to the magnetic anomalies
$\,a_{\mu,e}=(g_{\mu,e}-2)/2$\,
of charged leptons, sufficiently small to be compatible with upper limits
from present experimental results and theoretical evaluations.


The prospects for detecting light dark matter particles in terrestrial (direct) dark matter searches,
with the present technology based on the measurement of nuclear recoil energy, do not appear bright, since these direct detection experiments
are only sensitive to sizeable masses above about 7 GeV \citep{edelweiss}, or 1 GeV (Cresst \citep{cresst}, MACHe3, Rosebud and Tokyo), and not to the light dark matter masses considered here.

\section{Conclusion}

Summarizing, the 511 keV emission from the Galactic bulge observed by SPI/INTEGRAL could be the signature of light ($\,\approx$ 1 -- 100 MeV) dark matter particles, with a preference for the lower end of this mass interval. Such particles could annihilate throughout the Galactic bulge into positrons (and electrons) which, after being stopped, themselves annihilate into 511 keV photons.
The link between the positron annihilation line from the central Galactic regions and dark matter, if real, would push INTEGRAL at the forefront of astroparticle research.
It is thus essential to further test this hypothesis, and to discard, if possible, competing ones using INTEGRAL itself as well as other means.
For this purpose, it is crucial to get a detailed map of the 511 keV emission
in the central region of the Galaxy.

\section*{Acknowledgments}

The authors would like to thank C. Boehm, D. Hooper, J. Silk, E. Parizot and R. Lehoucq for illuminating discussions.


\end{document}